\documentclass[10pt]{iopart}
\usepackage[utf8]{inputenc}
\usepackage{graphicx}
\usepackage{epstopdf}
\usepackage{color}

\usepackage{iopams}  
\begin{document}

\title[Dynamics of liquid bridges]{Dynamics of two-dimensional liquid bridges}

\author{Rodrigo C. V. Coelho$^{1,2}$, Lu\'{\i}s A. R. G. Cordeiro$^{1,2}$, Rodrigo B. Gazola$^{1,2}$ and Paulo I. C. Teixeira$^{1,3}$}


\address{$^1$Centro de F\'{\i}sica Te\'{o}rica e Computacional, 
Faculdade de Ci\^{e}ncias, Universidade de Lisboa, 1749-016 Lisboa, Portugal \\ 
$^2$Departamento de F\'{\i}sica, Faculdade de Ci\^{e}ncias,
Universidade de Lisboa, 1749-016 Lisboa, Portugal \\
$^3$ISEL -- Instituto Superior de Engenharia de Lisboa,
Instituto Polit\'{e}cnico de Lisboa, 
Rua Conselheiro Em\'{\i}dio Navarro 1, 1959-007 Lisboa, Portugal}

\ead{rcvcoelho@fc.ul.pt and piteixeira@fc.ul.pt}
\vspace{10pt}

\begin{abstract}
We have simulated the motion of a single vertical, two-dimensional liquid 
bridge spanning the gap between two flat, horizontal solid substrates of 
given wettabilities, using a multicomponent pseudopotential 
lattice Boltzmann method. For this simple 
geometry, the Young–Laplace equation can be solved (quasi-)analytically to 
yield the equilibrium bridge shape under gravity, which provides a check 
on the validity of the numerical method. In steady-state conditions, we 
calculate the drag force exerted by the moving bridge on the confining 
substrates as a function of its velocity, for different contact angles and 
Bond numbers. We also study how the bridge deforms as it moves, as 
parametrized by the changes in the advancing and receding contact angles 
at the substrates relative to their equilibrium values. Finally, starting 
from a bridge within the range of contact angles and Bond numbers in which 
it can exist at equilibrium, we investigate how fast it must move in order 
to break up.
\end{abstract}

%
%
%
%
\ioptwocol
%


\section{Introduction}

\label{sec-intro}

A liquid bridge, also called capillary bridge, is a wall or column of liquid 
connecting two bodies, which may be solid surfaces (flat or curved), particles,
or other liquids \cite{Israelachvili:2011}. Liquid bridges are relevant in many
contexts, such as sand art \cite{sandart}; atomic-force 
microscopy in high-humidity environments \cite{AFM}; soldering \cite{solder};
the testing of weakly-adhesive solid surfaces \cite{weakly}; in lungs, where 
they may close small airways and impair gas exchange \cite{lungs}; the wet 
adhesion of insects and tree frogs \cite{frogs}; the feeding of shore birds 
\cite{birds}; the spontaneous filling of porous materials \cite{porous};
or as tools for contact angle measurements \cite{Nagy:2019}.

In an earlier paper \cite{Teixeira_2019}, we investigated the stability of 
two-dimensional (2D) liquid bridges under gravity: a slab of liquid between 
two flat, unbounded horizontal substrates at which the contact angles are 
fixed. Because of the very simple geometry, we were able to compute 
analytically the range of gap widths (as described by a Bond number ${\rm Bo}$) 
for which the liquid bridge can exist, as well as the conditions for the 
occurrence of necks, bulges and inflection points on its surface, for given 
contact angles at the top and bottom substrates. In addition, the shapes of 
these bridges could be computed by straightforward numerical integration.

At least as important as equilibrium bridges are moving bridges in some of
the above situations \cite{Howell:2000}, as well as in microfluidic devices
\cite{labonchip} or in the treatment of certain pulmonary 
diseases \cite{lung}, which requires liquid plugs to be mobilised, and 
eventually broken, in order to re-open the airways.
In particular, it would be useful to know how easily a bridge
(also called a `liquid plug') slides along a gap, i.e, what force must be 
applied to make it move at a certain velocity, as a function of gap width and 
the contact angles at the substrates. Intuitively, one might expect this force
to be smaller if the liquid does not wet the substrates. One other question that
may be asked is, how hard does one have to push in order to break a bridge that 
is a stable equilibrium state? Here we address these problems using a popular 
numerical method for the simulation of multicomponent flows.

This paper is organized as follows: in section~\ref{sec-method} we describe our
general simulation method (\ref{sec-method-lBm}), how we implement two-phase 
flows (\ref{sec-method-internal}), tune the wettabilities of substrates
(\ref{sec-method-wetting}) and compute the drag force exerted by the moving 
liquid bridge on the solid substrates (\ref{sec-method-drag}). Our results are
presented in section~\ref{sec-results}: first we validate our method by
comparing the shapes it yields for static bridges with the semi-analytical 
solutions of the Young-Laplace equation~\cite{Teixeira_2019} 
(\ref{sec-results-static}); then we establish how the drag force 
(\ref{sec-results-drag}) and contact angles (\ref{sec-results-wetting}) scale
with the bridge velocity, for different combinations of Bond numbers and 
substrate wettabilities; finally, we investigate the stability of many types of
bridges (i.e., the maximum velocity needed to break them) 
(\ref{sec-results-breakup}). We conclude in section~\ref{sec-concl}.

\section{Method}

\label{sec-method}

\subsection{Lattice-Boltzmann method}

\label{sec-method-lBm}

We simulate the liquid bridges using a multicomponent lattice Boltzmann 
method based on the Shan-Chen model, also known as pseudopotential 
model~\cite{PhysRevE.47.1815, kruger2016lattice}, sandwiched between two flat, 
horizontal solid substrates a distance $L_y$ apart. Let us denote the two 
components as $A$, the liquid bridge component, and $B$, the surrounding fluid, 
or, generically as $\sigma$ (where $\bar\sigma$ is the other component). Each 
component has its own distribution function $f_i^{(\sigma)}$ and is evolved 
independently according to the Boltzmann-BGK equation:
\begin{eqnarray}
  &&f_i^{(\sigma)}(\mathbf{x}+\mathbf{c}_i \Delta t, t+\Delta t) 
- f_i^{(\sigma)}(\mathbf{x}, t) = \nonumber \\
  && -\frac{\Delta t}{\tau^{(\sigma)}}\left[f_i^{(\sigma)}(\mathbf{x}, t)
-f_i^{eq(\sigma)}(\mathbf{x}, t)\right] + S_i^{(\sigma)}(\mathbf{x}, t)\Delta t,
\end{eqnarray}
where $\mathbf{c}_i$ is the $i$-th velocity vector of the D3Q19 lattice, 
$\Delta t$ is the time step, $\tau$ is the relaxation time, which controls 
the kinematic viscosity $\nu=(\tau-1/2)/3$, and $S_i$ is the forcing term. 
We set $\tau=1.2$ in our simulations. Notice that we use a 
more generic 3D lattice, but we set $L_z=1$ with periodic boundary conditions in the $z$-direction, which is equivalent to simulating 
a 2D system~\cite{kruger2016lattice}. The equilibrium distribution function is given by
\begin{eqnarray}
 && {f^{eq}_{i}}^{(\sigma)} = \nonumber \\
 &&\rho^{(\sigma)} w_i \left[ 1+ \frac{\mathbf{c}_i\cdot\mathbf{u}^{eq}}{c_s^2} 
+ \frac{(\mathbf{c}_i\cdot{\mathbf{u}^{eq}})^2}{2c_s^4} 
- \frac{(\mathbf{u}^{eq})^2}{2c_s^2}  \right],
\label{feq-eq}
\end{eqnarray}
where $\rho^{(\sigma)}$ is the density of component $\sigma$, $w_i$  are 
the weights of the D3Q19 lattice ($w_0 = 1/3$ for $\vert \mathbf{c}\vert^2=0$, 
$w_s = 1/18$ for $\vert \mathbf{c}\vert^2=1$ and $w_l = 1/36$ for 
$\vert \mathbf{c}\vert^2=2$) and $c_s=1/\sqrt{3}$ is the speed of sound in this 
lattice. Equation~(\ref{feq-eq}) is the second-order expansion 
of the Maxwell-Boltzmann distribution in Hermite 
polynomials~\cite{kruger2016lattice,COELHO2018144}.
The two fluids interact in two ways: through a common macroscopic 
velocity, and via the internal forces. The common macroscopic velocity 
that enters the equilibrium distribution is
\begin{eqnarray}
\mathbf{u}^{eq} = \frac{1}{\rho} \sum_\sigma 
\left( \sum_i f_i^{(\sigma)}\mathbf{c}_i 
+ \frac{\mathbf{F}^{(\sigma)}\Delta t}{2}\right),
\label{eq:velocity}
\end{eqnarray}
where
\begin{eqnarray}
\rho = \sum_\sigma \rho^{(\sigma)}
\end{eqnarray}
is the total density of the fluid. The total force acting on each component 
$\mathbf{F}^{(\sigma)}$ is a sum of the external force applied to the fluid and 
internal forces which are responsible for fluid segregation. The external 
force $\mathbf{F}^{\tiny{\mbox{ext}}}$ on each component is 
$\mathbf{F}^{\tiny{\mbox{ext}},(\sigma)}=\rho^{(\sigma)}\mathbf{F}^{\tiny{\mbox{ext}}}/\rho$.
The forces are implemented in the distribution function space using the 
Guo forcing term~\cite{PhysRevE.65.046308}:
\begin{eqnarray}
S_i ^{(\sigma)} =&& w_i\left(1-\frac{\Delta t}{2\tau^{(\sigma)}} \right) 
\nonumber \\
&& \times \left[ \frac{\mathbf{c}_i}{c_s^2}
\left(1+\frac{\mathbf{c}_i\cdot \mathbf{u}^{eq}}{c_s^2}\right) 
- \frac{\mathbf{u}^{eq}}{c_s^2} \right]\cdot \mathbf{F}^{(\sigma)} .
  \label{source-eq}
\end{eqnarray}
The dimensional quantities in this paper are given in lattice units, in which 
the density $\rho$, the time step $\Delta t$ and the lattice spacing $\Delta x$ 
are all unity. The lattice Boltzmann method described here recovers the Navier-Stokes and continuity equations in the macroscopic limit~\cite{kruger2016lattice}.

\subsection{Internal forces}

\label{sec-method-internal}

The pseudopotential models consider internal forces which are responsible for 
the segregation between the two components. These forces are calculated using 
pseudopotentials that depend on the local density of each component. The 
inter-component force reads:
\begin{eqnarray}
&&\mathbf{F}^{\tiny{\mbox{inter}}, (\sigma)}(\mathbf{x}) = \nonumber \\ 
&& - \rho^{(\sigma)}(\mathbf{x})\, G_{\sigma \bar \sigma} \sum _i w_i 
\,\rho^{(\bar \sigma)}(\mathbf{x}+\mathbf{c}_i\Delta t) \,\mathbf{c}_i \,\Delta t ,
\label{inter-force-eq}
\end{eqnarray} 
where $G_{\sigma \bar \sigma}$ controls the force strength and needs to be above a 
certain threshold in order to drive segregation. We set 
$G_{\sigma \bar \sigma}=G_{\bar\sigma \sigma}\equiv G_{AB} = 1.3$ in this work, in which 
the positive sign leads to a repulsive force. 

Equation~(\ref{inter-force-eq}) 
is sufficient to simulate two immiscible fluids. However, because we also want 
to investigate the effect of gravity on the liquid bridges, we need
the two components to have different densities. For this purpose, we also 
consider an intra-component cohesive force acting only on component $A$:
\begin{eqnarray}
&&\mathbf{F}^{\tiny{\mbox{intra}}, (A)}(\mathbf{x}) = \nonumber \\ 
&& - \psi^{(A)}(\mathbf{x})\, G_{AA} \sum _i w_i \,
\psi^{(A)}(\mathbf{x}+\mathbf{c}_i\Delta t) \,\mathbf{c}_i \,\Delta t ,
\label{intra-force-eq}
\end{eqnarray} 
where $G_{AA}$ is the strength of the intra-component force and we set it to 
$G_{AA}=-2$. The minus sign means the force is attractive, and therefore
$\rho^{(A)} > \rho^{(B)}$. 
The pseudopotential is $\psi^{(A)} = \rho_0 [1-\exp(-\rho^{(A)}/\rho_0 )]$, 
where we set the reference density as $\rho_0=1.5$.
The resulting equation of state is
\begin{eqnarray}
  p =&& c_s^2 \rho^{(A)} + c_s^2 \rho^{(B)} 
+  G_{AB}c_s^2 \Delta t^2 \rho^{(A)}\rho^{(B)}\nonumber \\
 &&+ \frac{1}{2} G_{AA} c_s^2 \Delta t^2 \psi^{(A)} \psi^{(A)},
\end{eqnarray}
where $p$ is the pressure.

\subsection{Wetting}

\label{sec-method-wetting}

We implement the wetting boundary condition as in 
Refs.~\cite{doi.org/10.1002/fld.4988, doi:10.1063/5.0080823}, which is summarized as follows. A field 
$\phi(\mathbf{x})$ is set to one at the solid nodes and zero at the fluid 
nodes. This scheme consists in setting a virtual solid density at the solid 
nodes which will control the contact angle. It is calculated using a weighted 
average of the density of the fluid neighbours:
\begin{eqnarray}
&& \tilde{\rho}^{(\sigma)}(\mathbf{x}) =\nonumber \\
&&\chi^{(\sigma)}\frac{\sum_i w_i \rho^{(\sigma)}(\mathbf{x}+\mathbf{c}_i\Delta t) 
(1-\phi(\mathbf{x}+\mathbf{c}_i\Delta t) ) }
{\sum_i w_i (1- \phi(\mathbf{x}+\mathbf{c}_i\Delta t)) }.
 \label{av-dens-eq}
\end{eqnarray}
The parameter $\chi$ is different for each component: $\chi^{(A)}=1-\xi$ and 
$\chi^{(B)}=1+\xi$, where $\xi$ is the wetting parameter that controls the 
contact angle. If $\xi<0$, the substrate attracts (wants to be wetted by)
component $A$, whereas if $\xi>0$, the substrate attracts (wants to be wetted 
by) component $B$. By abuse of language we shall refer to these two situations 
as `hydrophilic' and  `hydrophobic' substrates, respectively, although neither
component need be water. The weights and velocity vectors are 
those of the D3Q19 lattice. The adhesion force is already accounted for in 
Eqs.~(\ref{inter-force-eq}) and~(\ref{intra-force-eq}) where the sums also 
extend over the solid nodes. Thus, this scheme treats fluid-fluid and 
fluid-solid interactions in the same way provided that the solid density is 
calculated as in equation~(\ref{av-dens-eq}). The no-slip boundary condition 
applies to the fluid at the solid nodes, which is implemented using the 
half-way bounce-back conditions.

\subsection{Drag force}

\label{sec-method-drag}

A moving bridge exerts a hydrodynamic drag force on the substrates, which is 
calculated using the moment exchange method~\cite{PhysRevE.65.041203}, as 
(assuming the plates to be at rest):
\begin{eqnarray}
 \mathbf{F}_s =  \frac{\Delta x ^3}{\Delta t} \sum_{\mathbf{x}_s, i} 
2\left[ {f_{i}}^{\ast, (A)}(\mathbf{x}_s,t) 
+ {f_{i}}^{\ast, (B)}(\mathbf{x}_s,t) \right] \mathbf{c}_i,
 \label{drag-force-eq}
\end{eqnarray}
where ${f_{i}}^{\ast, (\sigma)}$ are the reflected distributions in the 
bounce-back boundary condition and $\mathbf{x}_s$ are the positions of 
the solid nodes.

\section{Results}

\label{sec-results}


\subsection{Static liquid bridges}
\label{sec-results-static}

In this section, we validate our hydrodynamic model by comparing its results
with predictions for static (equilibrium) liquid bridges~\cite{Teixeira_2019}.
For the simple 2D geometry under study, the Young-Laplace equation can be 
solved quasi-analytically to yield the equilibrium bridge shape under gravity. 
In \cite{Teixeira_2019} it was assumed that the bridge 
is immersed in air, i.e., 
that the density of the surrounding medium is much lower than that of the 
bridge and can be neglected. However, it is straightforward to show that 
exactly the same equation, and hence the same solution, is obtained if 
the Bond number, which measures the relative importance of gravity and 
surface tension, is redefined as 
\begin{eqnarray}
{\rm Bo} = \frac{\Delta \rho\,  F^{\tiny{\mbox{ext}}}_y \, L_Y^2}{\gamma},
\end{eqnarray}
where $\Delta\rho$ is the density difference between the bridge and the 
surrounding medium, and $F^{\tiny{\mbox{ext}}}$ is the force density (or acceleration)
 applied in the 
vertical direction (perpendicular to the substrates). The shape of the 
right-hand surface bounding the bridge is then~\cite{Teixeira_2019}
\begin{eqnarray}
\label{defxfinb}
&&x^\prime(y^\prime)=
x^\prime(0)-\nonumber \\
&&\int_0^{y^\prime} \frac{[-\cos\theta_T y^{\prime\prime}
+(1 - y^{\prime\prime}) \left( \cos \theta_B +
\frac{\rm Bo}{2} y^{\prime\prime}\right)]dy^{\prime\prime}}
{\sqrt{1-\left[-\cos\theta_T y^{\prime\prime}
+( 1 - y^{\prime\prime}) \left( \cos\theta_B + \frac{\rm Bo}{2}
y^{\prime\prime} \right)\right]^2 }},
\end{eqnarray}
where $x^{\prime}=x/L_Y$ and $y^{\prime}=y/L_Y$ are scaled coordinates, and 
$\theta_T$ and $\theta_B$ are the contact angles at the top and bottom 
substrates, respectively.  
One important consequence of equation (\ref{defxfinb}) is that, for given 
$\theta_T$ and $\theta_B$, the bridge can only exist if ${\rm Bo}$ lies 
between zero and
\begin{eqnarray}
\label{new2roots}
{\rm Bo}_{max} &=& 2\left(2-\cos\theta_B+\cos\theta_T\right) \nonumber \\
&&+
4\sqrt{\left(1-\cos\theta_B\right)\left(1+\cos\theta_T\right)}.
\end{eqnarray}
This is plotted in figure 2 of Ref.~\cite{Teixeira_2019}.

In the simulations, we initialize the densities of the 
two components as follows: inside the liquid bridge $\rho^{(A)} = 4$ and 
$\rho^{(B)} = 0.001$, and outside $\rho^{(A)} = 0.001$ and $\rho^{(B)} = 2$;
the density difference is thus $\Delta \rho \approx 2$. 
Except where stated otherwise, the simulation box dimensions are 
$L_x \times L_Y=1024 \times 64$ and the two initially flat interfaces which 
bound the liquid bridges are placed at $x_1=3L_x/8$ and $x_2=5L_x/8$. Using 
the Laplace test, we obtain for the surface tension between the two fluids with 
the internal forces described in the previous section 
$\gamma = 0.342 \pm 0.001$. 

\begin{figure}[h]
\center
\includegraphics[width=\linewidth]{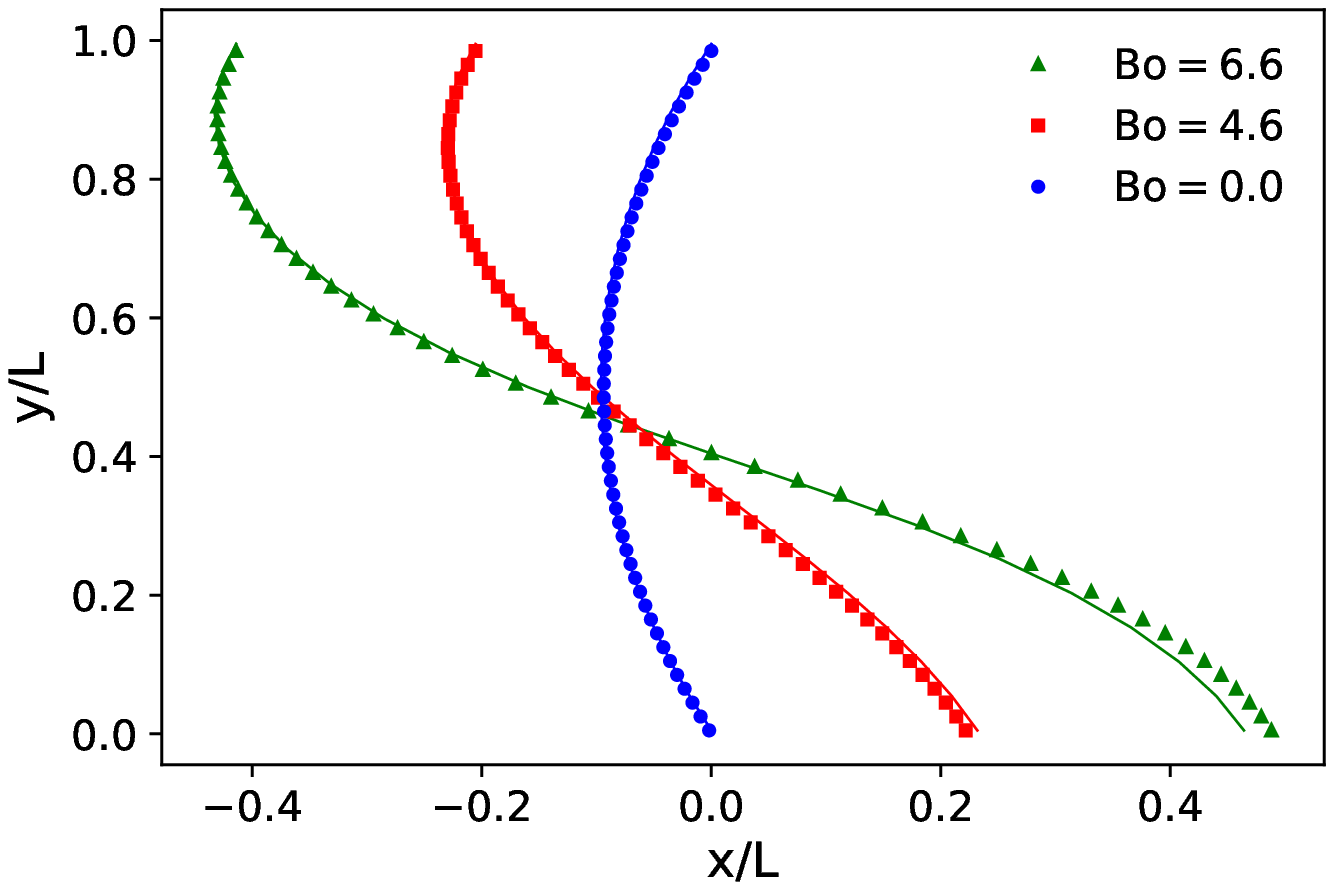}
\includegraphics[width=\linewidth]{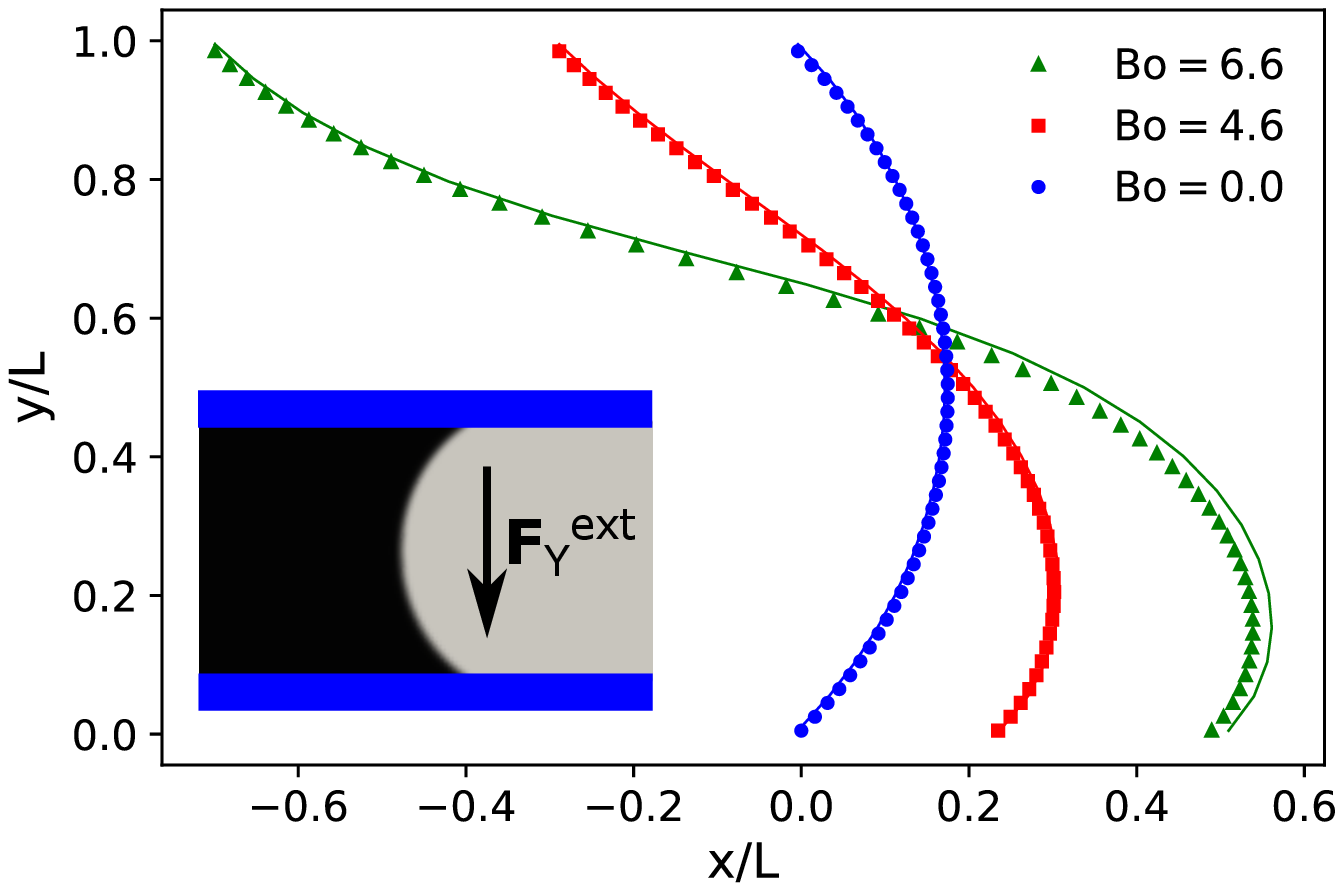}
\caption{Interface shape for $\theta\approx 51^\circ$ (top) and $112^\circ$
(bottom) and three different Bond numbers. Symbols are simulation results, 
solid lines are the semi-analytical predictions for each Bond number. 
The inset shows a sketch of the system. Here $L=L_Y=200$.}
\label{interface-fig}
\end{figure}

We start by comparing the shape of the liquid bridges obtained analytically 
and numerically. Figure~\ref{interface-fig} shows 
this comparison for two liquid contact angles $\theta=\theta_T=\theta_B$, 
corresponding to fairly hydrophilic ($\theta \approx 51^\circ$) and fairly 
hydrophobic ($\theta\approx 112^\circ$) substrates, and for each of these 
three different Bond numbers. 
For ${\rm Bo}=0$, the bridge surfaces -- the interfaces between $A$-rich 
and $B$-rich liquids -- are circular arcs; they become more top-down asymmetric
as ${\rm Bo}$ is increased. For each contact angle, the largest Bond number 
considered is close to the maximum for which a bridge can exist. We conclude 
that our numerical model faithfully reproduces the shape of a static bridge,
except when the Bond number approaches its maximum value.

\begin{figure}[h]
\center
\includegraphics[width=\linewidth]{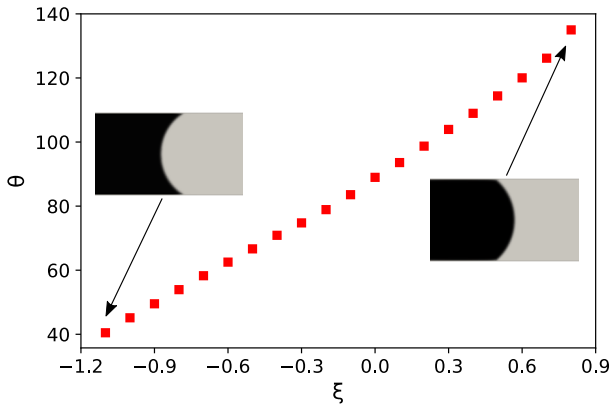}
\caption{Mapping between the wetting parameter $\xi$ and the contact angle 
$\theta$ of the liquid bridge (in degrees). The insets show two examples of bridge shape 
for a hydrophilic and a hydrophobic substrate with wetting parameter and 
contact angle indicated by the arrows.}
\label{contact-angle-fig}
\end{figure}

In figure~\ref{contact-angle-fig} we plot the contact angle $\theta$ of 
the static liquid bridges versus the wetting parameter $\xi$,
for ${\rm Bo}=0$. This curve provides a mapping between $\xi$ 
and $\theta$ under these conditions. Note that $\xi$ controls a material 
property of the solid surface and not the angle itself. Thus, the angle may 
vary for the same $\xi$ depending on Bond number, as will be discussed in 
the next sections.    
\begin{figure*}[h]
\center
\includegraphics[width=\linewidth]{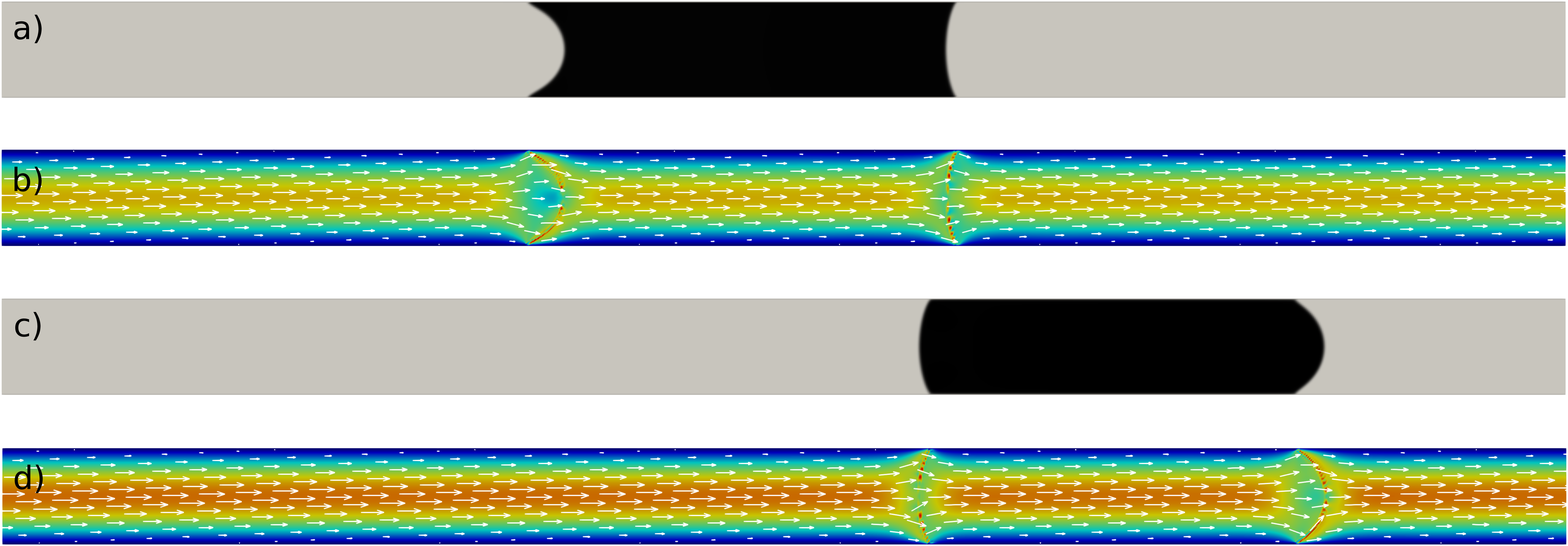}
\caption{Example of two liquid bridges moving due to a force 
$F^{\tiny{\mbox{ext}}}=10^{-5}$ lattice units between 
(a) two hydrophilic substrates, 
with $\xi=-0.8$, and (c) two hydrophobic substrates, with $\xi=0.7$. Panels 
(b) and (d) show the velocity fields corresponding to panels (a) and (c) 
respectively, where the colours denote the velocity magnitude (red for fast
and blue for slow) and the arrows indicate its direction. 
Note that the velocity is approximately the same for both fluids and that they are incompressible.}
\label{examples1-fig}
\end{figure*}

\subsection{Force on substrates}
\label{sec-results-drag}

We now analyze the fluid friction forces exerted on the substrates by a 
moving bridge. In order to make our results generic, we express them in terms 
of two non-dimensional quantities. The capillary number, ${\rm Ca}$, quantifies 
the relative importance of the viscous forces over the surface tension forces. 
In our simulations, since the surface tension is constant, ${\rm Ca}$ will be 
a measure of the bridge velocity. We define it as: 
\begin{eqnarray}
{\rm Ca} = \frac{\langle\rho\rangle \langle u \rangle \nu}{\gamma},
\end{eqnarray}
where $\langle \rho \rangle$ is the average density in the simulation domain, 
and $\langle u \rangle$ is the average velocity of the fluid in the 
$x$-direction, which, as we 
have checked, is the same as the interface velocity of the liquid bridge in 
the steady state. The other non-dimensional quantity is the drag coefficient,
${\rm Cd}$, which is a measure of the drag, or frictional, force exerted by the
liquid on the substrates:
\begin{eqnarray}
{\rm Cd} = \frac{F^{\tiny{\mbox{drag}}}}{\langle\rho\rangle\langle u \rangle^2 L_x}.
\end{eqnarray}
By dimensional analysis, the drag force on the substrates is of the form 
$F^{\tiny{\mbox{drag}}} \propto \langle \rho \rangle \langle u \rangle \nu L_x$.
Note that this is similar to Stokes' law for a sphere immersed in a viscous 
fluid, with the sphere radius replaced by $L_x$ and a proportionality constant 
equal to $6\pi$. For the liquid bridges, the proportionality constant 
depends on the contact angle. We expect that the drag force will be smaller for 
hydrophobic substrates than for hydrophilic ones. Therefore
\begin{eqnarray}
{\rm Cd} \sim \frac{\langle \rho \rangle \nu^2}{\gamma} {\rm Ca^{-1}}. 
\end{eqnarray}

We initialize the liquid bridges as described 
in section~\ref{sec-results-static} 
and apply an external force in the direction $x$ parallel to the substrates. 
After an initial transient, whose duration depends on the contact angles and 
the magnitude of the applied force, a steady-state is established in which the 
bridge moves with a constant velocity. This requires that the drag force on 
the substrates must have equal magnitude to the force applied
on the bridge, which provides a check on the consistency of our results.
Figure~\ref{examples1-fig} shows two examples of moving liquid bridges, 
between hydrophilic or hydrophobic substrates. The interfaces are set far 
enough apart that they do not interact with each other and a Poiseuille flow 
develops inside and outside the bridges. Only close to the interfaces are 
there disruptions to the velocity field. This occurs because 
the two fluids are immiscible, and therefore there is no relative velocity in 
the direction normal to the interface of the liquid bridge.

\begin{figure}[h]
\center
\includegraphics[width=\linewidth]{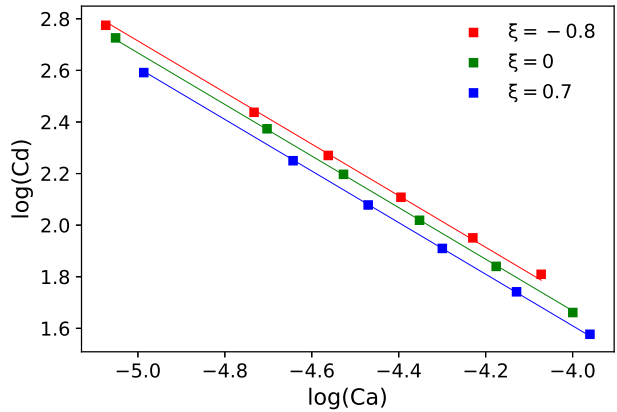}
\caption{Plot of the drag coefficient ${\rm Cd}$ {\it vs} capillary number 
${\rm Ca}$ for three wetting parameters $\xi$. The squares are simulation 
results, the solid lines are linear fits of the form $f(x)=-x +b$.}
\label{cd-ca-fig}
\end{figure}

In figure~\ref{cd-ca-fig}, the expected power law 
${\rm Cd}\propto {\rm Ca}^{-1}$
is confirmed, where the linear fits assume a slope of $-1$. We note that the 
hydrophilic substrate ($\xi=-0.8$) results in a greater drag force on the 
plates than the hydrophobic substrate ($\xi=0.7$) with the neutral wetting 
substrate ($\xi=0$) lying in between. 

\subsection{Dynamic contact angle}

\label{sec-results-wetting}

In this section, we measure the changes in contact angle when the bridge is 
moving, with respect to the bridge at rest. We consider three cases: same 
wetting parameter at both top, $\xi_T$, and bottom, $\xi_B$, substrates and 
${\rm Bo}=0$; different wetting parameters at top and bottom and ${\rm Bo}=0$; 
and same wetting parameter at top and bottom but with ${\rm Bo}\ne 0$. We point
out that the flow of liquid bridges is laminar: the Reynolds number, defined 
as ${\rm Re} = \langle u\rangle L_Y/\nu$, varies between $1.1$ and $3.1$.

\subsubsection{Identical substrates, no gravity.}
\label{equal-xi-sec}

\begin{figure}[h]
\center
\includegraphics[width=\linewidth]{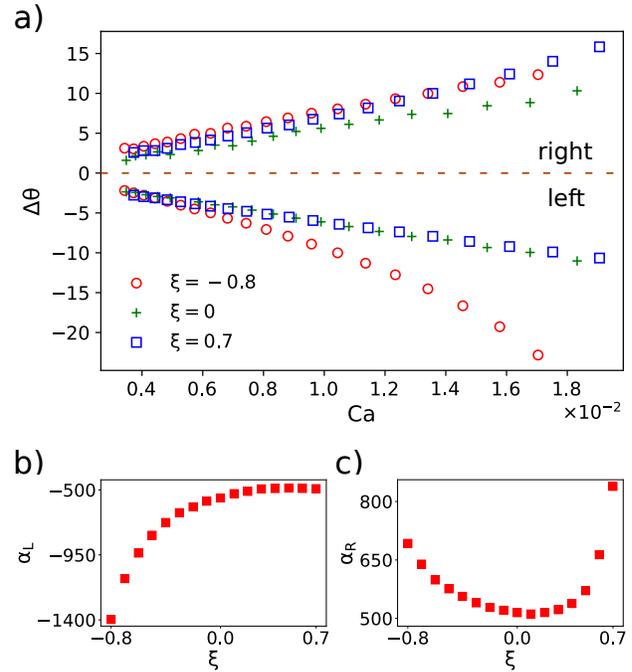}
\caption{(a) Change in contact angle $\Delta \theta$ relative to equilibrium
{\it vs} capillary number ${\rm Ca}$ for three wetting parameters $\xi$. 
$\Delta \theta >0$ for the advancing interface (on the right) and 
$\Delta \theta <0$ for the receding interface (on the left). The contact angles 
are the same at the top and bottom substrates. The slopes of the linear fits of
$\Delta \theta$ {\it vs} ${\rm Ca}$ are plotted in (b) and (c) for the receding 
(left) and advancing (right) angles, respectively. }
\label{slope1-fig}
\end{figure}

As illustrated in figure~\ref{examples1-fig}, the contact angles become 
different from those of the bridges at rest. Additionally, the angles of the 
advancing interface (on the right) increase while those of the retreating 
interface (on the left) decrease, for both hydrophilic and hydrophobic 
substrates. The top and bottom contact angles are the same if ${\rm Bo}=0$ 
and $\xi_T = \xi_B$ as here. In figure~\ref{slope1-fig}(a), we show how the 
angle changes as a function of ${\rm Ca}$ (varying velocities) for three 
different wetting parameters. We note that the slopes of $\Delta\theta$ for 
the advancing contact angles are always positive, but they might vary 
slightly with the wetting parameter and the slopes of 
$\Delta \theta$ for the receding contact angles are negative. Since 
$\Delta \theta \times {\rm Ca}$ is almost linear (except for receding angles
with $\xi=-0.8$), we use the slope of a linear 
fit $\Delta\theta(x)=\alpha\,{\rm Ca}$ as a measure of the angles' 
dependence on the wetting parameter. These slopes are shown in 
Figs.~\ref{slope1-fig}(b) and (c) for receding ($\alpha_L$) and advancing 
($\alpha_R$) angles, respectively. (The fit of $\Delta\theta$ for receding 
angles with $\xi=-0.8$ did not use the four points with largest ${\it Ca}$).
Slopes $\alpha_L$ vary more strongly with 
$\xi$ than $\alpha_R$. We note that $\vert \alpha_L \vert$ is larger
for smaller equilibrium contact angles. This happens because the receding 
angles tend to decrease and this is easier if they are already small due to 
the tendency of the interface to become more curved for higher velocities. 
Curiously, $\alpha_R$ has a minimum close to $\xi=0$. This means that it is 
more difficult to change the interface shape when it is flat 
($\theta=90^\circ$).

\subsubsection{Different substrates, no gravity.}

\begin{figure}[h]
\center
\includegraphics[width=\linewidth]{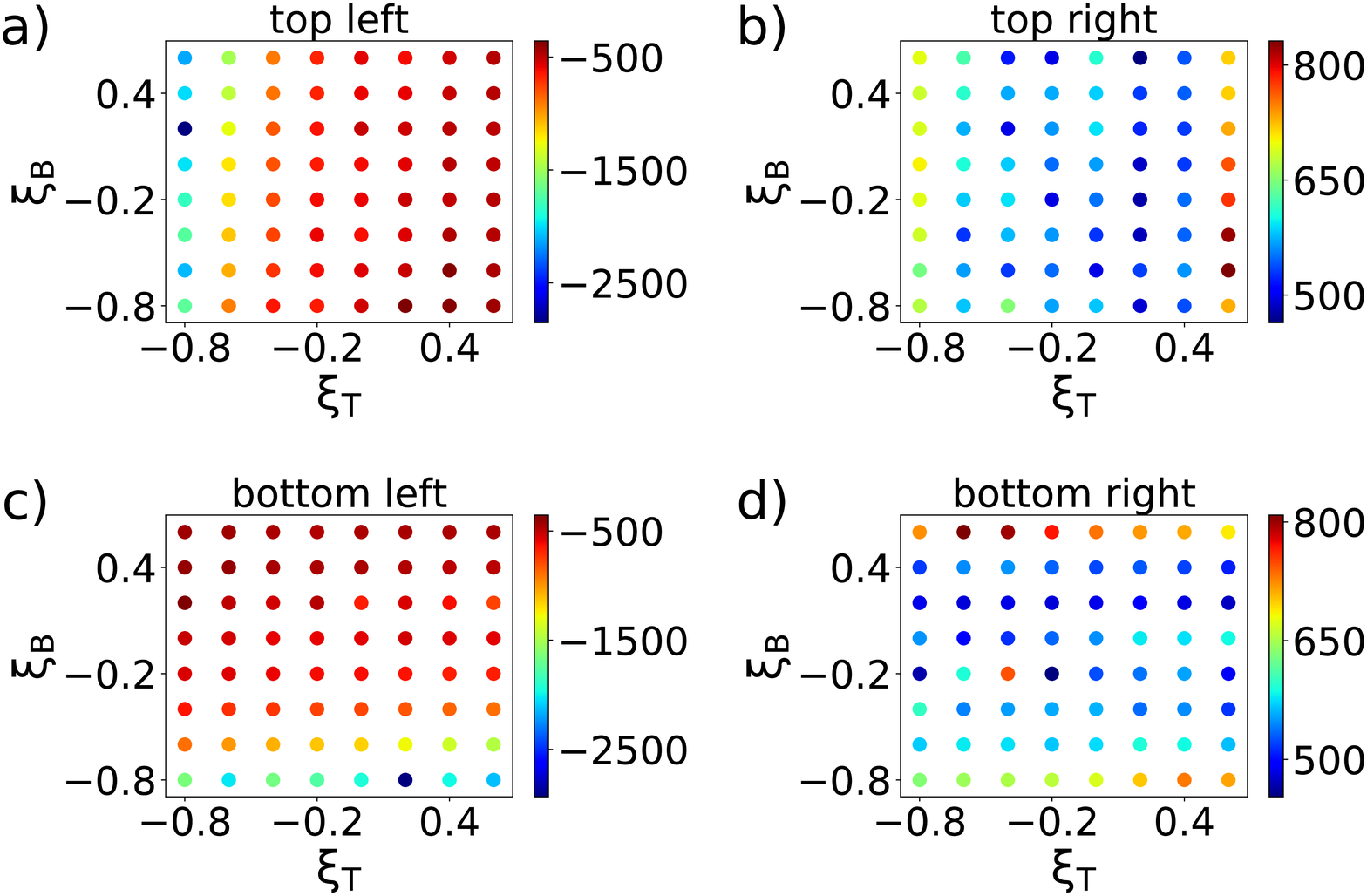}
\caption{Slopes of the linear fits of $\Delta \theta$ {\it vs} ${\rm Ca}$  
for the four contact angles of the liquid bridge with different wetting 
parameters at top and bottom, $\xi_T$ and $\xi_B$ respectively. }
\label{slopes2-fig}
\end{figure}

Here we consider different wetting parameters at the top and bottom substrates, 
$\xi_T$ and $\xi_B$ respectively. We proceed as in section~\ref{equal-xi-sec} 
and apply an external force in the direction parallel to the flow. Then we 
relate the change $\Delta \theta$ to the capillary number ${\rm Ca}$ 
and measure the slope of the linear fit for each combination of $\xi_T$ and 
$\xi_B$, see in figure~\ref{slopes2-fig}. Because the bridge is moving and 
the top and bottom wettabilities are different, all four contact angles are
different  and their slopes are shown separately. For the contact angles on 
the right (advancing), the slopes vary over a narrower range, and therefore
the errors of the linear fits and other numerical errors in contact angle 
measurements are more visible. These figures 
have a symmetry: the bottom angle on each side (left and right) for a given 
combination $\xi_T=\xi_1$ and $\xi_B=\xi_2$ is the same as the top angle on the 
same side with exchanged parameters $\xi_T=\xi_2$ and $\xi_B=\xi_1$. This does 
not hold for ${\rm Bo\ne 0}$ since the force in the $y$ direction breaks the 
top-bottom symmetry. For the angles on the left, which vary over a wider 
range, the slopes of the fits are only weakly affected by the wetting on the 
bottom substrate and vice-versa. Hydrophilic substrates ($\xi<0$) yield larger 
absolute slopes on the left-hand side. For the angles on the right, the slopes 
are smaller close to neutral wetting ($\xi=0$).  

\subsubsection{Identical substrates, ${\rm Bo} \ne 0$.}

\begin{figure}[h]
\center
\includegraphics[width=\linewidth]{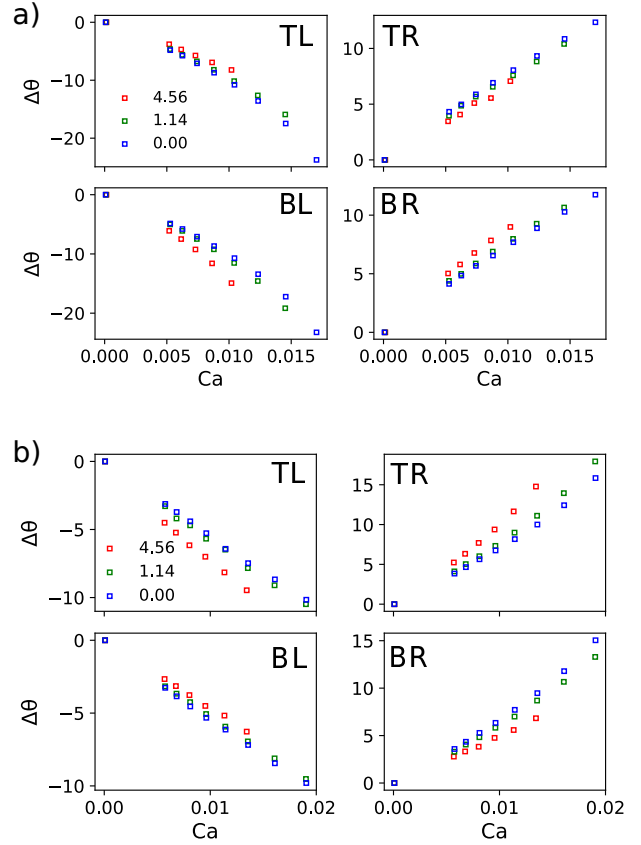}
\caption{Changes in contact angles $\Delta \theta$ {\it vs} capillary number 
${\rm Ca}$ for different Bond numbers (see legend: top left (TL), top right 
(TR), bottom left (BL), bottom right (BR). The wetting parameter is the same 
on the top and bottom substrates: (a) $\xi=-0.8$ and (b) $\xi=0.7$.}
\label{bo-fig}
\end{figure}

Now we consider bridges with the same wetting parameter at top and bottom but 
with an external force (e.g., gravity) perpendicular to the plates. We measure 
the contact angle change $\Delta\theta$ as a function of the capillary number 
$Ca$ as before. Figure~\ref{bo-fig} shows the variations of the four angles for 
three different Bond numbers and two wetting parameters. The external force in 
the $y$ direction increases some angles and decreases others. Those angles that 
decrease on the left interface vary more strongly with capillary number  
than those on the right interface. For instance, the top left (TL) angle 
increases for a hydrophilic top substrate as ${\rm Bo}$ increases. Thus, its 
variation with ${\rm Ca}$ is smaller for larger ${\rm Bo}$ as shown in 
figure~\ref{bo-fig}(a). The opposite happens for a hydrophobic top substrate: 
as the TL angle decreases with ${\rm Bo}$, its variation with ${\rm Ca}$ is 
greater for larger ${\rm Bo}$. The same analysis holds for 
the other three angles.

\subsection{Breaking bridges}

\label{sec-results-breakup}

In this section we examine the maximum capillary number 
${\rm Ca}_{\tiny{\mbox{max}}}$ for which the liquid bridge is stable and does 
not break. Thus, using the same setup as in the previous sections, we 
simulate bridges with larger external forces applied in the $x$ direction 
to observe bridge rupture. We say the bridge has broken if, after a long 
enough time ($t=5\times 10^5$), there is no possible path inside the liquid 
bridge component connecting the top and bottom substrates. 

\begin{figure}[h]
\center
\includegraphics[width=\linewidth]{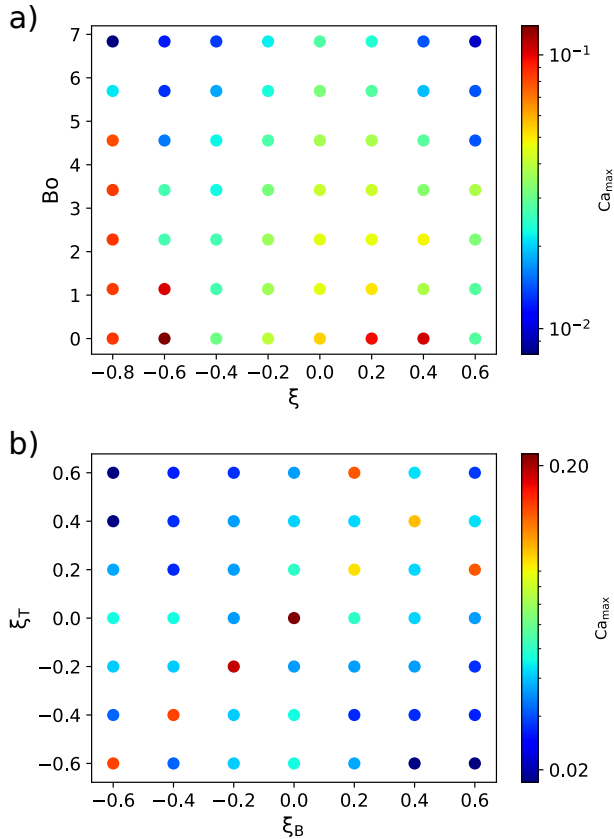}
\caption{Maximum capillary number ${\rm Ca}_{\tiny{\mbox{max}}}$ 
for which the liquid bridge does not break. (a) Same wetting parameter 
$\xi$ at top and bottom substrates and varying Bond number ${\rm Bo}$. 
(b) ${\rm Bo}=0$ and different wetting parameters on top and bottom.}
\label{break-fig}
\end{figure}

Figure~\ref{break-fig}(a) shows ${\rm Ca}_{\tiny{\mbox{max}}}$ for liquid bridges 
between identical substrates and different Bond numbers. We notice 
that bridges with wetting parameters around $\xi=0$ (neutral wetting) are, in 
general, more stable for any ${\rm Bo}$. This is because 
the interface is straight for $\xi=0$ and increasing ${\rm Bo}$ does not 
much affect its shape. Bridges between hydrophilic substrates 
are also more resilient, presumably owing to the larger liquid-solid contact
area that provides greater adhesion. Increasing ${\rm Bo}$ makes bridges more 
unstable, especially far from $\xi=0$. To have an idea of what
these numbers mean in physical units, let us consider a liquid bridge 
made of a commercially available aqueous surfactant solution (Pustefix, 
Germany) in air, for which $\Delta\rho\sim 10^3$~kg~m$^{-3}$, 
$\gamma\sim 28$~mJ~m$^{-2}$ and $\nu\sim 10^{-6}$~m$^2$~s$^{-1}$. 
For $\xi=-0.6$ and ${\rm Bo}=1.1$, which corresponds to $L_y=1.8$~mm, the 
maximum capillary number is ${\rm Ca}_{\tiny{\mbox{max}}}\approx 0.1$, which gives
a bridge velocity of $3$~m/s. In this case, the drag coefficient is 
${\rm Cd}= 0.95$, whence the drag force on the plates is $250$~N.
For the same $\xi$ and ${\rm Bo}=6.8$, which corresponds to $L_y=4.6$ mm, 
${\rm Ca}_{\tiny{\mbox{max}}}\approx 0.012$, which gives a velocity of 
$0.36$~m/s. The drag coefficient in this case is ${\rm Cd}= 9.1$ and 
the corresponding drag force is $87$~N. Finally, note that, according to our 
earlier work \cite{Teixeira_2019}, the maximum span of a liquid bridge at 
equilibrium is four capillary lengths.

In figure~\ref{break-fig}(b), we plot ${\rm Ca}_{\tiny{\mbox{max}}}$ for bridges 
with ${\rm Bo}=0$, but different wetting parameters at top and bottom. Bridges 
with $\xi_T \neq \xi_B$ are more unstable, which effect is stronger close to 
the extremes, with $\xi_T=0.6$ and $\xi_B=-0.6$ and $\xi_T=-0.6$ 
and $\xi_B=0.6$.  

It is interesting to note that, in all cases considered here, the smallest
capillary number for which a bridge breaks is less than that needed to break 
a thin liquid lamella, ${\rm Ca}\sim 0.26$~\cite{Cantatrev}.

\section{Conclusions}

\label{sec-concl}

We have simulated the motion of a two-dimensional liquid bridge sandwiched 
between flat horizontal substrates of tunable wettability, using the lattice
Boltzmann method. This was validated by comparison with quasi-analytic results
for equilibrium (static) bridges~\cite{Teixeira_2019}, and found to be reliable
except for Bond numbers very close to the maximum above which an equilibrium 
bridge breaks.

For steady-state bridges moving between identical substrates, 
${\rm Cd}\sim{\rm Ca}^{-1}$ as required by dimensional analysis. However
the prefactor depends on the liquid  contact angles, being larger for 
hydrophilic than for hydrophobic substrates. This accords with the intuitive 
expectation that bridges of a given volume should have a larger contact area 
with a hydrophilic than with a hydrophobic substrate, thus leading to higher 
drag in the former case than in the latter. The effect is, however, not very 
pronounced: drag reduction due to hydrophobicity is at most about $20\%$. 

Moving bridges deform from their equilibrium shapes. We quantified this by
measuring the deviations of contact angles from their equilibrium values.
For not-too-fast moving bridges these deviations are proportional to the 
capillary number with slopes that are always negative on the receding side of
a bridge and positive on the advancing side, but exhibit a complex dependence 
on the contact angles.

Finally, we investigated bridge stability by mapping the largest capillary
number for which a bridge will not break, for given contact angles at the top 
and bottom substrates and given Bond number. Interestingly, the most stable 
bridges appear to be those with equilibrium contact angles around $90^\circ$.
Stability (i.e, breakup deferred to larger ${\rm Ca}$) is also favoured by
hydrophilicity and small Bond numbers.

One possible area of future research would be the detailed breakup of bridges, 
either as a result of increasing their velocity (i.e., ${\rm Ca}$) or substrate
separation (i.e., ${\rm Bo}$). One possible extension would be to non-Newtonian
fluids, which occur in many industrial and biological contexts.

\section*{Acknowledgments}
We acknowledge financial support from the Portuguese Foundation for Science 
and Technology (FCT) under the contracts: EXPL/FIS-MAC/0406/2021, PTDC/FIS-MAC/5689/2020, 
UIDB/00618/2020 and UIDP/00618/2020.
\\\\

 \bibliography{refs} 
 \bibliographystyle{unsrt}

\end{document}